\documentstyle[12pt]{article}

\renewcommand{\theequation}{\thesection.\arabic{equation}}

\newlength{\extraspace}
\newlength{\extraspaces}
\newcounter{dummy}

\newcommand{\baa}{
\addtocounter{equation}{1}
\setcounter{dummy}{\value{equation}}
\setcounter{equation}{0}
\renewcommand{\theequation}{\thesection.\arabic{dummy}\alph{equation}}
\begin{eqnarray}
\addtolength{\abovedisplayskip}{\extraspaces}
\addtolength{\belowdisplayskip}{\extraspaces}
\addtolength{\abovedisplayshortskip}{\extraspace}
\addtolength{\belowdisplayshortskip}{\extraspace}}

\newcommand{\eaa}{
\end{eqnarray}
\setcounter{equation}{\value{dummy}}
\renewcommand{\theequation}{\thesection.\arabic{equation}}}

\newcommand{\be}{\begin{equation}
\addtolength{\abovedisplayskip}{\extraspaces}
\addtolength{\belowdisplayskip}{\extraspaces}
\addtolength{\abovedisplayshortskip}{\extraspace}
\addtolength{\belowdisplayshortskip}{\extraspace}}
\newcommand{\ee}{\end{equation}}

\newcommand{\ba}{\begin{eqnarray}
\addtolength{\abovedisplayskip}{\extraspaces}
\addtolength{\belowdisplayskip}{\extraspaces}
\addtolength{\abovedisplayshortskip}{\extraspace}
\addtolength{\belowdisplayshortskip}{\extraspace}}
\newcommand{\ea}{\end{eqnarray}}

\newcommand{\bd}{\begin{displaymath}
\addtolength{\abovedisplayskip}{\extraspaces}
\addtolength{\belowdisplayskip}{\extraspaces}
\addtolength{\abovedisplayshortskip}{\extraspace}
\addtolength{\belowdisplayshortskip}{\extraspace}}
\newcommand{\ed}{\end{displaymath}}

\newcommand{\ban}{\begin{eqnarray*}
\addtolength{\abovedisplayskip}{\extraspaces}
\addtolength{\belowdisplayskip}{\extraspaces}
\addtolength{\abovedisplayshortskip}{\extraspace}
\addtolength{\belowdisplayshortskip}{\extraspace}}
\newcommand{\ean}{\end{eqnarray*}}

\newcommand{\newsection}[1]{
\vspace{15mm}
\pagebreak[3]
\addtocounter{section}{1}
\setcounter{equation}{0}
\setcounter{subsection}{0}
\setcounter{footnote}{0}
\begin{center}
{\Large \thesection. #1}
\end{center}
\nopagebreak
\medskip
\nopagebreak}

\newcommand{\newsubsection}[1]{
\vspace{1cm}
\pagebreak[3]

\addtocounter{subsection}{1}
\noindent{ \sc \thesubsection. #1}
\nopagebreak
\vspace{2mm}
\nopagebreak}

\newcommand{\nonu}{\nonumber \\[.5mm]}

\newcommand{\deel}[2]{{\textstyle{#1 \over #2}}}

\newtheorem{thm}{Theorem}
\newtheorem{lmm}{Lemma}
\newtheorem{exam}{Example}
\newcommand{\bt}{\begin{thm}}
\newcommand{\et}{\end{thm}}
\newcommand{\bl}{\begin{lmm}}
\newcommand{\el}{\end{lmm}}
\newcommand{\bex}{\begin{exam}}
\newcommand{\eex}{\end{exam}}
\newcommand{\hj}{\hat{J}}

\def\inbar{\,\vrule height1.5ex width.4pt depth0pt}
\font\rms=cmr12 at 12pt
\def\ce{\relax\ifmmode\mathchoice
{\hbox{$\inbar\kern-.3em{\rm C}$}}
{\hbox{$\inbar\kern-.3em{\rm C}$}}
{\lower.9pt\hbox{\rms $\inbar\kern-.3em{\rm C}$}}
{\lower1.2pt\hbox{\rms $\inbar\kern-.3em{\rm C}$}}
\else{$\inbar\kern-.3em{\rm C}$}\fi}
\font\cmss=cmss12 \font\cmsss=cmss12 at 12pt
\def\ze{\relax\ifmmode\mathchoice
{\hbox{\cmss Z\kern-.4em Z}}{\hbox{\cmss Z\kern-.4em Z}}
{\lower.9pt\hbox{\cmsss Z\kern-.4em Z}}
{\lower1.2pt\hbox{\cmsss Z\kern-.4em Z}}\else{\cmss Z\kern-.4em Z}\fi}

\newcommand{\dif}{\partial}

\newcommand{\actie}[1]{\deel{1}{2\pi}\int d^2z \, }

\newcommand{\mats}[9]{\left( \begin{array}{ccc}
				#1 & #2 & #3 \\
				#4 & #5 & #6 \\
				#7 & #8 & #9
                             \end{array} \right) }

\newcommand{\mat}[4]{\left( \begin{array}{cc}
				#1 & #2 \\
				#3 & #4
                             \end{array} \right) }

\begin{document}
\addtolength{\baselineskip}{.7mm}

\thispagestyle{empty}
\begin{flushright}
{\sc THU}-93/05\\
{\sc ITFA}-02-93
\end{flushright}
\vspace{1.5cm}
\setcounter{footnote}{2}
\begin{center}
{\LARGE\sc{The relation between quantum
W algebras and Lie algebras}}\\[0.8cm]

\sc{Jan de Boer\footnote{e-mail: deboer@ruunts.fys.ruu.nl}}
\\[1.5mm]
{\it Institute for Theoretical Physics\\[0.7mm]
Princetonplein 5\\[0.7mm]
P.O. Box 80.006\\[0.7mm]
3508 TA Utrecht\\[6.5mm]}
\sc{ Tjark Tjin \footnote{e-mail: tjin@phys.uva.nl}}
\\[1.5mm]
{\it Institute for Theoretical Physics\\[0.7mm]
Valckenierstraat 65\\[0.7mm]
1018 XE Amsterdam\\[0.7mm]
The Netherlands\\[10mm]}

\message{baselineskip=\the\baselineskip}

{\sc Abstract}\\[0.5cm]
\end{center}

\noindent
{\footnotesize \baselineskip 12pt
By quantizing the generalized Drinfeld-Sokolov reduction scheme for
arbitrary $sl_2$ embeddings we show that a large set $\cal W$ of quantum
W algebras can be viewed as (BRST) cohomologies of affine Lie algebras.
The set $\cal W$ contains many known $W$ algebras such as $W_N$ and
$W_3^{(2)}$. Our formalism yields a completely algorithmic method for
calculating the W algebra generators and their operator product
expansions, replacing the cumbersome construction of W algebras as
commutants of screening operators. By generalizing and quantizing
the Miura transformation we show that any $W$ algebra in $\cal W$
can be embedded into the universal enveloping algebra of a semisimple
affine Lie algebra which is, up to shifts in level, isomorphic
to a subalgebra of the original affine algebra. Therefore {\em any}
realization of this semisimple affine Lie algebra leads to a realization
of the $W$ algebra. In particular, one obtains in this way a general and
explicit method for constructing the free field realizations and Fock
resolutions for all algebras in $\cal W$.  Some examples are explicitly
worked out.

\message{\baselineskip=\the\baselineskip}}

\vfill

\newpage

\newsection{Introduction}
$W$ algebras were introduced by Zamolodchikov as  a new
ingredient in the classification program of conformal field (CFT)
theories \cite{Za} (for a recent review see \cite{BoSc}).
As is well known such a classification would correspond to a
classification of all possible perturbative groundstates of string
theory. However CFT and W algebras have been shown to be related
to several other areas of research as well such as integrable systems,
2D critical phenomena and the quantum Hall effect. $W$ symmetries are
therefore an interesting new development in theoretical physics and it
is the purpose of this paper to provide a step towards understanding
their meaning and structure.

The point of view that we shall develop in this paper is that the theory
of $W$ algebras is closely related to the theory of Lie algebras and
Lie groups. The construction of $W$ algebras as Casimir algebras
(with as a special case the Sugawara construction), and the
coset construction \cite{GKO,BBSS} are examples of such a relation,
but unfortunately these have some serious drawbacks. We therefore
take the Hamiltonian or Drinfeld-Sokolov (DS) \cite{DS,BFFWO}
reduction perspective
which for classical $W$ algebras has been extremely successful.

The DS reduction approach starts
with the observation that certain Poisson algebras encountered in the
theory of integrable evolution equations can be considered to be
classical versions of the $W$ algebras first constructed by Zamolodchikov
\cite{BM}. Drinfeld and Sokolov had already shown that these Poisson
algebras are reductions of Kirillov Poisson structures on the duals
of affine Lie algebras thus providing a relation between Lie algebras
and $W$ algebras. A first attempt to
quantize the classical $W$ algebras found by Drinfeld-Sokolov reduction
(nowadays called $W_N$ algebras) was made in \cite{FL}. There the Miura
transformation was used to realize the generators of the classical $W_N$
algebra in terms of classical free fields. The algebra was then
quantized by making the free fields into quantum free fields and
normal ordening the expressions of the $W$ generators. In general this
is not a valid quantization procedure however since it is by no means
clear that the algebra of quantum $W$ generators will close. In fact it
only closes in certain cases \cite{BoSc}
(which are ofcourse the cases that were
studied in \cite{FL}).

Since DS reduction is in essence Hamiltonian reduction in infinitely many
dimensions it is possible to apply the techniques of BRST quantization
in order to quantize the classical $W_N$ algebras.
This was first done in \cite{beroog}
for the special case of the Virasoro algebra and the $W_N$ case was solved
by Feigin and Frenkel \cite{FeFr}.

Even though $W_N$ algebras have an appealing description as
BRST cohomologies of affine Lie algebras the quantum DS
method is still rather limited since $W_N$ algebras are by
far not the only $W$ algebras.
The quantum DS reduction leading to the by now well known $W_3^{(2)}$
algebra \cite{P,B} was however the first indication that DS reduction
can be generalized to include many other $W$ algebras.

In \cite{BTV} it has been shown that to every $sl_2$
embedding into the simple Lie algebra underlying the affine algebra
there is associated a generalized classical DS
reduction of this affine algebra leading to
a $W$ algebra. The fact that one considers $sl_2$ embeddings is
closely related to the fact that one wants the reduced algebra
to be an extended conformal algebra (i.e. it must contain the Virasoro
algebra as a subalgebra and the other generators must be primary
fields w.r.t. this Virasoro algebra).  Since the number of
inequivalent $sl_2$
embeddings into $sl_n$ is equal to the number of partitions of the
number $n$ the set of $W$ algebras that can be obtained by
DS reduction increased drastically.  The $W_N$ algebras turn out to be
associated to the so called 'principal' $sl_2$ embeddings. The
Polyakov-Bershadsky algebra $W_3^{(2)}$ is associated to the
only nonprincipal $sl_2$ embedding into $sl_3$.

The reductions considered in \cite{BTV} are classical
and it is the purpose of this paper to quantize them.
The usual formalism developed in \cite{beroog,FeFr} constructs the
$W$ algebra as the commutant of certain screening operators which is
rather difficult to generalize to arbitrary $sl_2$ embeddings.
The main reason for this is that it is difficult to find
a complete set of generators of this algebra for arbitrary $sl_2$
embeddings (one has to make use of character formulas to check if
one has obtained all generators. These characters are however not known
in advance).
Also it makes use of free field realizations of the
original affine Lie algebra which means that one obtains, in the end,
the $W$ algebra in its free field form. If the $W$ algebra
has affine subalgebras this will therefore get obscured by the
not very transparant free field form.
In this paper we therefore use
the formalism that was developed in \cite{BT} to quantize finite $W$
algebras. It turns out that this formalism still works, with some
modifications, in the infinite dimensional case considered here.

Let us now give an outline of the paper. In section 2 we quantize
the generalized DS reductions of \cite{BTV}. This is done by
the same spectral sequence calculation that was used in \cite{BT}.
We also introduce the quantum Miura transformation for arbitrary
$sl_2$ embeddings and show how to obtain free field realizations for
arbitrary $W$ algebras. In section 3 we briefly discuss the conformal
properties of the quantum $W$ algebras obtained in section 2. Furthermore
a general formula for the central charges of the $W$ algebras in terms
of the level of the affine Lie algebra and the defining vector of the
$sl_2$ embedding is given. In the last section we consider some examples
in order to illustrate the general procedure. We end the paper with
some comments and open problems.

\newsection{Quantization}
Let $\{t_a\}$ be a basis of the Lie algebra $\bar{g} \equiv sl_n$.
The affine Lie
algebra $g$ associated to $\bar{g}$ is the span of $\{J^a_n\}$
and the central element $K$. The commutation relations are given by
\be
[J^a_n,J^b_m]  =  f^{ab}_cJ^c_{n+m}+ng^{ab}K\delta_{n+m,0} \;\; ; \;\;
[K,J^a_n]  =  0  \label{comm}
\ee
where $g^{ab}$ is the inverse of
$g_{ab}=\mbox{Tr}(t_at_b)$ and $[t_a,t_b]=f_{ab}^ct_c$.
As usual we use $g_{ab}$ to raise and lower indices.
Let ${\cal U}_kg$ $({k \in \bf C})$  be the universal enveloping algebra
of $g$ quotiented by the ideal generated by $K-k$.

It was shown in \cite{BTV} that one can associate
to every $sl_2$ embedding
into $\bar{g}$ a Drinfeld-Sokolov reduction of $g$ leading to a classical
$W$ algebra. We shall now quantize these algebras. Let
$\{t_0,t_+,t_-\}$ be an $sl_2$ subalgebra of $\bar{g}$ then one can decompose
$\bar{g}$ into eigenspaces of the operator $ad_{t_0}$
\be \label{grading}
\bar{g}=\bigoplus_{k\in \frac{1}{2} {\bf N}} \bar{g}_k \label{een}
\ee
where $\bar{g}_k=\{x \in \bar{g} \mid [t_0,x]=k\, x\}$.
This defines a gradation of $\bar{g}$ which is in general
half integer. However, it
was shown in \cite{FORTW,BT} that in those cases where the grading contains
half integers one can replace it by an integer grading which in the end
leads to the same Drinfeld-Sokolov reduction.
This is done by replacing $t_0$ by a certain element $\delta$ of the Cartan
subalgebra which has the property that the
grading w.r.t. the operator $ad_{\delta}$ is an integer grading (for some
basic facts on $sl_2$ embeddings and the explicit construction of $\delta$
given an $sl_2$ embedding see the appendix).
Without loss of generality
we can therefore assume that the gradation (\ref{een})
is integral. The algebra $\bar{g}$ now admits a triangular
decomposition into a direct sum of a
negative grade piece, a zero grade piece and a
positive grade piece denoted by $\bar{g}_-,\bar{g}_0$ and $\bar{g}_+$
respectively.

Under the adjoint action of the $sl_2$ subalgebra $\{t_0,t_{\pm}\}$, $\bar{g}$
decomposes into a direct sum of $sl_2$ multiplets. Let us choose the basis
$\{t_a\}$ such that all elements $t_a$ are basis vectors of some $sl_2$
multiplet.
Ofcourse this means that all $t_a$ are homogeneous
w.r.t. the gradation.
{}From now on we let latin indices $a,b,\ldots$ run over the entire basis
of $\bar{g}$, Greek indices $\alpha, \beta, \ldots$ over the  basis of
$\bar{g}_+$
and barred Greek indices $\bar{\alpha}, \bar{\beta},\ldots$ over the
basis of $\bar{g}_0 \oplus \bar{g}_-$ (i.e. $\lambda^{\alpha}t_{\alpha} +
\lambda^{\bar{\alpha}}t_{\bar{\alpha}}=\lambda^at_a)$.

We now come to the constraints.  Since the $sl_2$ subalgebra
$\{t_0,t_{\pm}\}$ is a triplet under its own adjoint action  there
must be some $\alpha_+$ such that
$t_+=A\, t_{\alpha_+}$.
Define the character $\chi$ of $g_+$  (where $g_+$ is the affinization
of $\bar{g}_+$) by putting $\chi (J^{\alpha}_n) = A\delta^{\alpha ,\alpha_+}
\delta_{n+1,0}$
The constraints one imposes are then $J^{\alpha}_n=\chi (J^{\alpha}_n)$.
These constraints are first class \cite{BTV}
for integral gradings which means that one can
use the BRST formalism. Thereto introduce the fermionic ghost variables
$c_{\alpha}^n,b^{\alpha}_n$ with ghost numbers 1 and $-1$
respectively and relations
$c_{\alpha}^nb^{\beta}_m+b^{\beta}_mc_{\alpha}^n =\delta_{\alpha \beta}
\delta_{n+m,0}$
The algebra generated by these ghost variables is the Clifford algebra
$Cl(g_+ \oplus g_+^*)$. As usual one then considers the algebra
$\Omega_k = {\cal U}_kg \otimes Cl(g_+ \oplus g_+^*)$.

For calculational purposes it is convenient (as is standard
practice in conformal field theory) to introduce the following 'basic fields'
$J^a(z)  =  \sum_n J^a_n \,z^{-n-1}; \; c_{\alpha}(z) =
\sum_n c^n_{\alpha}\,z^{-n} ; \; b^{\alpha}(z) =
\sum_n b^{\alpha}_n \,z^{-n-1}$.
It is well known that the commutation relations in $\Omega_k$ can then be
expressed in terms of the operator product expansions (OPE)
\begin{eqnarray}
J^a(z) J^b(w) & = & \frac{kg^{ab}}{(z-w)^2} + \frac{f^{ab}_c}{z-w}J^c(w)
+ \ldots \\
c_{\alpha}(z)b^{\beta}(w) & = & \frac{\delta^{\beta}_{\alpha}}{z-w}
\end{eqnarray}
Now inductively define the {\em algebra of fields} $F(\Omega_k )$
as follows: $J^a(z),c_{\alpha}(z),b^{\alpha}(z)$
are elements of $F(\Omega_k )$
with 'conformal dimensions' $\Delta =1,0,1$ respectively;
if $A(z),B(z)\in F(\Omega_k )$ then $\alpha A(z)+\beta B(z) \in F
(\Omega_k )$;
if $A(z)$ is an element of $F(\Omega_k)$
of conformal dimension $\Delta$ then $\frac{dA}{dz}(z)$ is also an
element of $F(\Omega_k )$ and has conformal dimension $\Delta+1$;
if $A(z),B(z)$ are elements of $F(\Omega_k )$ of conformal dimensions
$\Delta_A$ and $\Delta_B$ respectively then
the normal ordened product $(AB)(z)\equiv A_-(z)B(z)\pm
B(z)A_+(z)$ (where one has the minus sign if $A$ and  $B$ are
fermionic) is also an element of
$F(\Omega_k )$ and has conformal dimension $\Delta_A + \Delta_B$.
Here $A_-(z)=\sum_{n\leq -\Delta_A}A_n\, z^{-n-\Delta_A}$ and $A_+(z)=
A(z)-A_-(z)$.
We say that $F(\Omega_k )$ is 'generated' by the basic fields. Note that
$F(\Omega_k ) \subset {\cal U}_kg[[z,z^{-1}]]$.
The algebra $F(\Omega_k)$ is graded by ghost number, i.e. $J^a(z)$,
$c_{\alpha}(z)$ and $b^{\alpha}(z)$ have degrees 0,1 and -1 respectively
and we have the decomposition
\be
F(\Omega_k)=\bigoplus_n F(\Omega_k)^{(n)}
\ee

The algebra of fields $F(\Omega_k)$ is not simply the set of
`words' in the fields that can be made using the rules given
above, there are also relations. If we denote the operator
product expansion of $A$ and $B$ by $A(z)B(w)\sim\sum_r\frac{
\{AB\}_r}{(z-w)^r}$, then the relations valid in
$F(\Omega_k)$ are \cite{BBSS}
\ba
(AB)(z)-(BA)(z)\equiv [A,B](z)
& = & \sum_{r>0} (-1)^{r+1}\frac{\partial^r}{r!}
\{AB\}_r \nonu
(A(BC))(z)-(B(AC))(z) & = & ([A,B]C)(z) \nonu
\dif(AB)(z) & = & (\dif A B)(z) + (A\dif B)(z). \label{rel}
\ea

The BRST operator is then \cite{KoSt} $D(.)=[d,.]$
where $d=\oint \frac{dz}{2\pi i}d(z)$ and
\be
d(z)=(J^{\alpha}(z)-\chi (J^{\alpha}(z)))c_{\alpha}(z)-\frac{1}{2}
f^{\alpha \beta}_{\gamma}(b^{\gamma}(c_{\alpha}c_{\beta}))(z)
\ee
$D$ is of degree 1 (i.e. $D(F(\Omega_k)^{(l)}) \subset F(\Omega_k)^{(l+1)})$
and $D^2=0$ which means that $F(\Omega_k)$ is a complex. One is then interested
in calculating the cohomology (or Hecke algebra) of this complex because
the zeroth cohomology is nothing but the quantization of the classical
$W$ algebra \cite{KoSt,FeFr}. This problem has been solved for the so called
'finite $W$ algebras' in \cite{BT}.

The first step is to split the BRST current into two pieces \cite{FeFr}
\begin{eqnarray}
d_0(z) & = &  -\chi (J^{\alpha}(z)) c_{\alpha}(z) \\
d_1(z) & = & J^{\alpha}(z) c_{\alpha}(z)-\frac{1}{2}
f^{\alpha \beta}_{\gamma} (b^{\gamma}(c_{\alpha}c_{\beta}))(z)
\end{eqnarray}
and to make $F(\Omega_k)$ into a double complex
$F(\Omega_k)= \bigoplus_{rs} F(\Omega_k)^{(r,s)}$
by assigning the following (bi)grades to its generators
\begin{eqnarray}
\mbox{deg}(J^a(z)) & = & (-k,k) \;\;\mbox{     if } \;t_a \in
\bar{g}_k \nonumber \\
\mbox{deg}(c_{\alpha}(z)) & = & (k,1-k) \;\;\mbox{     if } \;
t_{\alpha} \in \bar{g}_k
\nonumber \\
\mbox{deg}(b^{\alpha}(z)) & = & (-k,k-1) \;\;\mbox{     if }\; t_{\alpha} \in
\bar{g}_{k}
\end{eqnarray}
The operators $D_0:F(\Omega_k)^{(r,s)}\rightarrow F(\Omega_k)^{(r+1,s)}$ and
$D_1:F(\Omega_k)^{(r,s)} \rightarrow F(\Omega_k)^{(r,s+1)}$ associated in the
obvious way to $d_0$ and $d_1$ satisfy $D_0^2=D_1^2=
D_0D_1+D_1D_0=0$ verifying that we have obtained a double comlex..

Let us now calculate the action of the operators $D_0$ and $D_1$ on
the generators of $F(\Omega_k)$. For this it is convenient to introduce
$\hat{J}^a(z)=J^a(z)+f^{a\beta}_{\gamma}(b^{\gamma}c_{\beta})(z)$.
One then finds by explicit calculation
\begin{eqnarray}
D_0(\hat{J}^a(z)) & = & -f^{a\beta}_{\gamma}\chi(J^{\gamma}(z))c_{\beta}(z)
\nonumber \\
D_0(c_{\alpha}(z)) & = & 0 \nonumber \\
D_0(b^{\alpha}(z)) & = & -\chi (J^{\alpha}(z)) \nonumber \\
D_1(\hat{J}^a(z)) & = & f^{\alpha a}_{\bar{\beta}}\hat{J}^{\bar{\beta}}(z)
c_{\alpha}(z)+kg^{a\alpha}\partial c_{\alpha}(z)-f^{\alpha
e}_{\beta} f^{\beta a}_{e} \partial
c_{\alpha}(z) \nonumber \\
D_1(c_{\alpha}(z)) & = & -\frac{1}{2}f^{\beta \gamma}_{\alpha}(
c_{\beta}c_{\gamma})(z) \nonumber \\
D_1(b^{\alpha}(z)) & = & \hat{J}^{\alpha}(z). \nonumber
\end{eqnarray}
{}From these formulas it immediately follows that $D(\hat{J}^{\alpha}(z))=0$
and $D(b^{\alpha}(z))=\hat{J}^{\alpha}(z)-\chi (J^{\alpha}(z))$. This means
that the subspace $F^{\alpha}(\Omega_k)$
of $F(\Omega_k)$ generated by $J^{\alpha}(z)$ and
$b^{\alpha}(z)$ is actually a subcomplex. The cohomology of this complex
can easily be calculated and one finds $H^*(F^{\alpha}(\Omega_k);D)={\bf C}$.
Note also that due to the Poincare-Birkhoff-Witt theorem
for field algebras (which follows immediately from the relations
(\ref{rel})) the normal ordening map
\be
(\ldots ):F_{red}(\Omega_k) \otimes \bigotimes_{\alpha}
F^{\alpha}(\Omega_k) \rightarrow
F(\Omega_k)
\ee
defined by $A_1(z)\otimes \ldots \otimes A_l(z)
\mapsto (A_1\ldots A_l)(z)$
(where we always use the convention $(ABC)(z)=(A(BC))(z)$)
is an isomorphism of vectorspaces.
Due to this and the fact that the BRST operator
acts as a derivation\footnote{This follows from the fact that
$D(X(w))=\{d(z)X(w)\}_1$, and from the following general identities
for operator product expansions:
$\{A\dif B\}_1=\dif \{AB \}_1$ and $\{A(BC)\}_1=(-1)^{AB}
(B\{AC\}_1)+(\{AB\}_1C)$.}
on $F(\Omega_k)$ we have
\begin{eqnarray}
H^*(F(\Omega_k);D) & \simeq & H^*(F_{red}(\Omega_k);D)\otimes
\bigotimes_{\alpha}H^*(F^{\alpha}(\Omega_k);D)\nonumber \\
& \simeq & H^*(F_{red}(\Omega_k);D)
\end{eqnarray}
Where in the first step we used a Kunneth like theorem given in \cite{BT}.

In order to calculate  $H^*(F_{red}(\Omega_k);D)$
one uses the fact that $F_{red}(\Omega_k)$ is actually a double complex
which makes calculation of the cohomology possible via a spectral
sequence argument \cite{spec,FeFr,BT}. The first term $E_1$ of the spectral
sequence is the $D_0$ cohomology of $F_{red}(\Omega_k)$.
Note that we can write $D_0(\hat{J}^{\bar{\alpha}}(z))=-\mbox{Tr}([t_+,
t^{\bar{\alpha}}]t^{\beta}c_{\beta}(z))$.
Therefore $D_1(\hat{J}^{\bar{\alpha}}(z))=0$ iff $t_{\bar{\alpha}}
\in \bar{g}_{lw}$ where $\bar{g}_{lw}$ is the set of elements of $\bar{g}$
that are annihilated by $ad_{t_-}$ (the lowest weight vectors of the
$sl_2$ multiplets) and where we used the fact \cite{BT} that $t_{\bar{\alpha}}
\in \mbox{Ker(ad}_{t_-})$ iff $t^{\bar{\alpha}}\in \mbox{Ker(ad}_{t_+})$.
It can also easily be seen that for all $\beta$
there  exists a linear combination
$a(\beta )_{\bar{\alpha}} \hat{J}^{\bar{\alpha}}(z)$ such that
$D_0(a(\beta )_{\bar{\alpha}}\hat{J}^{\bar{\alpha}}(z))=c_{\beta}(z)$.
{}From this it follows \cite{BT}
that purely on the level of vectorspaces we have
\be
H^n(F_{red}(\Omega_k);D_0) \simeq F_{lw}(\Omega_k)\delta_{k,0}
\ee
where $F_{lw}(\Omega_k)$ is the subspace of $F(\Omega_k)$ generated by
the fields $\{J^{\bar{\alpha}}(z)\}_{t_{\bar{\alpha}}\in \bar{g}_{lw}}$.
Since the only cohomology that is nonzero is  of degree 0 the
spectral sequence abuts at the first term, i.e. $E_{\infty}=E_1$ and we
find the end result
\be
H^n(F_{red}(\Omega_k);D) \simeq F_{lw}(\Omega_k)\delta_{k,0}
\ee

Having calculated the BRST cohomology at the level of vector spaces
one now can construct the cohomology (or $W$ algebra)
generators and their OPEs  via a
procedure called the tic tac toe
construction \cite{BoTu}.
Consider a generator $\hat{J}^{\bar{\alpha}}(z)$ of degree $(p,-p)$
of the field algebra
$F_{lw}(\Omega_k)$ (i.e. $t_{\bar{\alpha}}\in \bar{g}_{lw}$) then
the generator of cohomology associated to this element
is given by
\be \label{tttc}
W^{\bar{\alpha}}(z)=\sum_{l=0}^p (-1)^l W^{\bar{\alpha}}_l(z)
\ee
where $W^{\bar{\alpha}}_0(z)\equiv J^{\bar{\alpha}}(z)$ and
$W^{\bar{\alpha}}_l(z)$ can be determined inductively by
\be
D_1(W^{\bar{\alpha}}_l(z))=D_0(W^{\bar{\alpha}}_{l+1}(z)) \label{ind}
\ee
It is easy to
check, using the fact that $D_0(J^{\bar{\alpha}}(z))=0$ for
$t_{\bar{\alpha}}\in \bar{g}_{lw}$ that indeed
$D(W^{\bar{\alpha}}(z))=0$.

The formalism presented above provides us with a completely algorithmic
procedure of calculating the $W$ algebra associated to a certain
$sl_2$ embedding: First determine the space $\bar{g}_{lw}$. Then take
a current $\hat{J}^{\bar{\alpha}}(z)$ with $t_{\bar{\alpha}} \in
\bar{g}_{lw}$ and inductively calculate the fields $W^{\bar{\alpha}}_l(z)$
using relations (\ref{ind}). The field (\ref{tttc}) is then the
corresponding W generator and the relations in the $W$ algebra are then
just the OPEs between the fields
$\{W^{\bar{\alpha}}(z)\}_{
t_{\bar{\alpha}}\in \bar{g}_{lw}}$
calculated using the OPEs in  $F(\Omega_k)$.

In principle this
algebra closes only modulo $D$-exact terms. But since we
computed the $D$ cohomology on a reduced comlex generated by
$\hj^{\bar{\alpha}}(z)$ and $c_{\alpha}(z)$, and this reduced complex
is zero at negative ghost number, there simply aren't any $D$
exact terms at ghost number zero. Thus the algebra generated by
$\{W^{\bar{\alpha}}(z)\}_{
t_{\bar{\alpha}}\in \bar{g}_{lw}}$ closes in itself.

As was shown in \cite{BT} for finite $W$ algebras,
the operator product algebra
generated by the fields $W^{\bar{\alpha}}(z)$ is isomorphic to the operator
product algebra generated by
their (bi)grade (0,0) components $W^{\bar{\alpha}}_p(z)$ (the proof in the
infinite dimensional case is completely analogous and will not be repeated
here).
The fields $W^{\bar{\alpha}}_p(z)$ are of course elements of the field
algebra generated by the currents $\{\hat{J}^{\bar{\alpha}}(z)\}_{
t_{\bar{\alpha}} \in \bar{g}_0}$. The relations (i.e. the OPEs)
satisfied by these currents are almost identical to the relations satisfied
by the unhatted currents
\be
\hat{J}^{\bar{\alpha}}(z)\hat{J}^{\bar{\beta}}(w)=\frac{kg^{\bar{\alpha}
\bar{\beta}}+ k^{\bar{\alpha}\bar{\beta}}}{(z-w)^2}+\frac{f^{\bar{\alpha}
\bar{\beta}}_{\bar{\gamma}}\hat{J}^{\bar{\gamma}}(w)}{z-w}+ \ldots
\label{hat}
\ee
where $k^{\bar{\alpha}\bar{\beta}}=f^{\bar{\alpha}\lambda}_{\gamma}
f^{\bar{\beta}\gamma}_{\lambda}$. Now, it is easy to see that
$\bar{g}_0$ is just a direct sum
of $sl_{p_j}$ and $u(1)$ algebras, i.e. forgetting for a moment about the
$u(1)$ algebras one can write
\be
\bar{g}_0 \sim \bigoplus_j sl_{p_j}
\ee
Within the $sl_{p_j}$ component of $\bar{g}_0$ we have the identity
\be
k^{\bar{\alpha}\bar{\beta}}=
g^{\bar{\alpha}\bar{\beta}}(h-h_j) \label{shift}
\ee
where $h$ is the dual coxeter number of $\bar{g}$ and $h_j$ is
the dual coxeter number of $sl_{p_j}$. We therefore find that the
field algebra generated by the currents $\{\hat{J}^{\bar{\alpha}}(z)\}_{
t_{\bar{\alpha}} \in \bar{g}_0}$, denoted from now on by
$\hat{F}_0$, is nothing but the field algebra
associated to a semisimple affine Lie algebra
the components of which are affine $sl_{p_j}$ and $u(1)$
Lie algebras. This semisimple affine
Lie algebra is not simply $g_0$ (whose field algebra is generated
by the unhatted currents) however
because in $g_0$ all components have the same level while in $\hat{F}_0$
the level varies from component to component as follows from
equation (\ref{shift}). This is just a result
of the ghost contributions $k^{\bar{\alpha}\bar{\beta}}$ in  the OPEs
of the hatted currents.

{}From the above we find that the  map
\be
W^{\bar{\alpha}}(z) \mapsto (-1)^p W^{\bar{\alpha}}_p(z)
\ee
is an embedding of the $W$ algebra into $\hat{F}_0$.
This provides a quantization
and generalization to arbitrary $sl_2$ embeddings of the well known
Miura map.
In \cite{FL} the standard Miura map for $W_N$
algebras was naively quantised by simply
normal ordening the classical expressions. This is known to work only
for certain algebras \cite{BoSc}. Our construction gives a rigorous
derivation of the quantum Miura transformations for arbitrary Kac-Moody
algebras and $sl_2$ embeddings (the generalized Miura transformations
for a certain special class of $sl_2$ embeddings were also recently
given in \cite{De}).

As a result of the generalized quantum Miura transformation {\em any}
representation or realization of $\hat{F}_0$ gives rise to a representation
or realization of the $W$ algebra. In particular one obtains a free field
realization of the $W$ algebra by choosing the Wakimoto free field realization
of $\hat{F}_0$. Given our formalism it is therefore straightforward to
construct
free field realizations for any $W$ algebra that can be obtained by
Drinfeld-Sokolov reduction.

\newsubsection{The Stress energy tensor}

It is possible to give a general expression for the
stress-energy tensor of a $W$ algebra related to an arbitrary
$sl_2$ embedding. For this purpose we write $t_0$ as
$t_0=s^at_a$, where the $s^a$ is only nonzero if $t_a$ lies in
the Cartan subalgebra. Furthermore, let $\delta_{\alpha}$ be the
eigenvalue of $ad_{t_0}$ acting on $t_{\alpha}$, thus
$[t_0,t_{\alpha}]=\delta_{\alpha}t_{\alpha}$. From this it is
easy to see that $\delta_{\alpha}=s_af^{\alpha a}_{\alpha}$.
Then the stress-energy tensor is
\be \label{t1}
T=\frac{1}{2(k+h)}\left(g_{a_0b_0}(\hj^{a_0}\hj^{b_0}) +
2 g_{b\alpha} \hj^b \chi(J^{\alpha}) - 2 ( k+h) s_a\dif\hj^a +
g_{b\alpha}f^{b\alpha}_e\dif\hj^e\right),
\ee
where the indices $a_0,b_0$ run only over $\bar{g}_0$, and $h$
is again the dual Coxeter number. By adding a $D$-exact term $D(R)$ to
(\ref{t1}), where
\be \label{RR}
R=\frac{1}{k+h}g_{b\alpha}(J^bJ^{\alpha})+\frac{1}{2(k+h)}
g_{e\alpha}f^{e\beta}_{\gamma}(b^{\alpha}(b^{\gamma}c_{\beta})),
\ee
we can rewrite it as
\be \label{t2}
T=\frac{1}{2(k+h)}g_{ab}(J^aJ^b)-s_a\dif J^a+(\delta_{\alpha}-1)
b^{\alpha}\dif c_{\alpha}+\delta_{\alpha}\dif b^{\alpha}
c_{\alpha},
\ee
which has the familiar form of improved Sugawara stress-energy
tensor plus the stress energy tensors of a set of free $b-c$
systems. The other generators of the $W$ algebra cannot in
general be written as the sum of a current piece plus a ghost
piece. Actually, (\ref{t2}) is precisely what one would expect
to get from a constrained WZNW model. Notice that
$\delta_{\alpha}$ is the degree of $t_{\alpha}$ with respect to
$t_0$, whereas $\alpha$ in (\ref{t2}) runs over $\bar{g}^+$
which was defined with respect to a new, different, integral
grading of the Lie algebra.

In terms of the level $k$ and the Cartan element of the
$sl_2$ embedding $t_0$ (called the 'defining vector'
since it determines the whole $sl_2$
subalgebra up to inner automorphisms)
the central charge of the $W$ algebra is given by \cite{FORTW}
\be \label{cc}
c(k;t_0)=r_{\bar{g}}-\frac{1}{2}\dim(\bar{g}_{\frac{1}{2}})-12\left|
\frac{\rho}{\sqrt{k+h}}-t_0\sqrt{k+h} \right|^2
\ee
where $r_{\bar{g}}$ is the rank of $\bar{g}$,
$\bar{g}_{\frac{1}{2}}$ is defined by (\ref{grading}), and
$\rho$ is half the sum of the positive roots, $\rho=\frac{1}{2}
\sum_{a\in\Delta^+} f^{b_0a}_{a} t_{b_0}$.

\newsection{Examples}
In this section we consider the three simplest cases of quantum
Drinfeld-Sokolov reduction, namely the Virasoro algebra, the Zamolodchikov
$W_3$ algebra and the so called Polyakov-Bershadsky algebra $W_3^{(2)}$.
For notational convenience we shall omit the
explicit $z$ dependence of the fields
where possible.

\newsubsection{The Virasoro algebra}

The Virasoro algebra is the simplest $W$ algebra and it is well
known to arize from the affine $sl_2$ KM algebra by
quantum Drinfeld-Sokolov reduction \cite{beroog}. It is the
$W$ algebra associated to the only nontrivial embedding of $sl_2$ into itself,
namely the identity map. We consider this example here to contrast our
methods to the ones used by Bershadsky and Ooguri.

Choose the following basis of
$sl_2$
\be \label{basis1}
J^at_a=\mat{J^2/2}{J^3}{J^1}{-J^2/2}.
\ee
where $t_0=-t_2, \; t_+=t_1$ and $t_-=t_3$.
The positive grade piece of the Lie algebra $\bar{g}=sl_2$ is generated by
$t_1$, and the constraint to be imposed is $J^1=1$. The BRST
current $d(z)$ is given simply by
\be \label{brst1}
d(z)=(J^1 c_1)-c_1.
\ee
The `hatted' currents are $\hj^1=J^1,\hj^2=J^2+2(b^1c_1)$ and
$\hj^3=J^3$. The
actions of $D_0$ and $D_1$ are given by
\be \label{d0d1}
\begin{array}{lclclcl}
D_0(\hj^2) & = & -2c_1 & \qquad & D_1(\hj^3) & = & (\hj^2c_2)+
(k+2)\dif c_1 \\
D_0(b^1) & = & -1 & \qquad & D_1(b^1) & = & \hj^1 .
\end{array}
\ee
On the other fields $D_0$ and $D_1$ vanish. From (\ref{d0d1}) it
is immediately clear that \\ $H(F_{red}(\Omega_k);D_0)$ is generated by
$W^3_0\equiv\hj^3$, in accordance with the general arguments in
section 2. To find the generator of the $D$-cohomology,
we apply the tic-tac-toe construction. We are looking for an
element $W^3_1(z)\in F_{red}(\Omega_k)$ such that
$D_0(W^3_1(z))=D_1(W^3_0)$. The strategy is to
write down the most general form of $W^3_1(z)$, and then to fix
the coefficients. In general, $W^{\bar{\alpha}}_l$ must satisfy
the following two requirements
\begin{enumerate}
\item if $W^{\bar{\alpha}}_l$ has bidegree $(-k,k)$, then
$W^{\bar{\alpha}}_{l+1}$ must have bidegree $(-k-1,k+1)$
\item if we define inductively the {\it weight} $h$
of a monomial in the
$J^{\bar{\alpha}}$ by $h(J^{\bar{\alpha}})=1-k$ if
$t_{\bar{\alpha}}\in g_k$,
$h((AB))=h(A)+h(B)$ and $h(\dif A)=h(A)+1$\footnote{
The weight $h$ is similar to the conformal weight, but not
always equal to it.
It is independent of the way in which the
$\hj$ are ordered.}, then $h(W^{\bar{
\alpha}}_l)=h(W^{\bar{\alpha}}_{l+1})$
\end{enumerate}
These two conditions guarantee that the most general form of
$W^{\bar{\alpha}}_l$ will contain only a finite number of
parameters, so that in a sense
the tic-tac-toe construction is a finite
algorithm. In the case under hand, the most general form of
$W^3_1$ is $a_1 (\hj^2\hj^2) + a_2 \dif \hj^2$, and the $D_0$ of
this equals $-4a_1(\hj^2c_1)-(4a_1+2a_2)\dif c_1$. Thus,
$a_1=-1/4$ and $a_2=-(k+1)/2$. Since $D_1(W^3_1)=0$, the
tic-tac-toeing stops here and the generator of the $D$
cohomology reads
\be \label{deft}
W^3=W^3_0-W^3_1=\hj^3+\frac{1}{4} (\hj^2\hj^2)
+\frac{(k+1)}{2}\dif\hj^2.
\ee
As one can easily verify, $T=W^3/(k+2)$ generates a Virasoro algebra
with central charge
\be
c(k)=13-6(k+2)-\frac{6}{k+2}
\ee
a result first found
by Bershadsky and Ooguri \cite{beroog}.

Let's now consider the quantum Miura transformation.
In the case at hand $\bar{g}_0$ is the Cartan subalgebra
spanned by $t_2$ and $\hat{F}_0$ is an affine $u(1)$ field algebra at
level $k+2$
generated by $\hat{J}^2$. Indeed defining the field $\partial \phi \equiv
\nu \hat{J}^2$ where $\nu = \sqrt{2(k+2)}$, it is easy to check that
\be
\partial \phi (z) \, \partial \phi (w) = \frac{1}{(z-w)^2}+\ldots
\ee
In terms of the field $\phi$ the (bi)grade (0,0)
piece of $T$ is given by
\be
T^{(0,0)}= \frac{1}{2}(\partial \phi \partial \phi) + \alpha_0
\partial^2 \phi   \label{V}
\ee
where $\alpha_0=\frac{1}{2}\nu-\frac{1}{\nu}$. This is the usual
expression for the Virasoro algebra in terms of a free field $\phi$.

\newsubsection{The Zamolodchikov $W_3$ algebra}

Having illustrated the construction in some detail for the
Virasoro algebra, we will now briefly discuss two other examples. We start
with the Zamolodchikov $W_3$ algebra \cite{Za}.
This algebra is associated to the so called 'principal' $sl_2$ embedding
in $sl_3$ \cite{BTV}. In terms of the following basis of $sl_3$
\be \label{basis3}
J^at_a=\mats{\frac{J^4}{6}+\frac{J^5}{2}}{\frac{1}{2}(J^6+J^7)}{
J^8}{\frac{1}{2}(J^2+J^3)}{-\frac{J^4}{3}}{\frac{1}{2}(-J^6+J^7)}{
J^1}{\frac{1}{2}(J^2-J^3)}{\frac{J^4}{6}-\frac{J^5}{2}}
\ee
the $sl_2$ subalgebra is given in this case by $t_+=4t_2,\; t_0=-2t_5$
and $t_-=2t_7$. The constraints are
therefore $J^1=J^3=0$ and
$J^2=1$ according to the general prescription. The BRST current
associated to these (first class) constraints reads
\be \label{brst3}
d(z)=J^1c_1+(J^2-1)c_2+J^3c_3+2(b^1(c_2c_3)).
\ee
The cohomology $H(F_{red}(\Omega_k);D_0)$ is generated by $\hj^7$ and
$\hj^8$ since $t_7$ and $t_8$ span $\bar{g}_{lw}$.
The tic-tac-toe construction gives as generators of $H(F(\Omega_k)
;D)$ the fields $W^7=W^7_0-W^7_1$ and $W^8=W^8_0-W^8_1+W^8_2$ where
\ba W^7_0 & = & \hj^7, \nonu
W^7_1 & = & -\frac{1}{6} (\hj^4\hj^4) - \frac{1}{2} (\hj^4\hj^5)
- 2(k+2)\dif\hj^5, \nonu
W^8_0 & = & \hj^8, \nonu
W^8_1 & = & -(\hj^5\hj^6)+\frac{1}{3}(\hj^4\hj^7)
-(k+2)\dif\hj^6, \nonu
W^8_2 & = & -\frac{1}{27}(\hj^4(\hj^4\hj^4)) + \frac{1}{3}
(\hj^4(\hj^5\hj^5)) + \frac{(k+2)}{3} (\hj^4\dif\hj^4) +
(k+2)(\hj^5\dif\hj^4) \nonu
& & +\frac{2(k+2)^2}{3}\dif^2\hj^4.
\label{ttt}
\ea
With some work, for instance by using the program for computing
OPE's by Thielemans \cite{thie}, one finds that
\ba
T & = & \frac{1}{2(k+3)}\; W^7, \nonu
W & = & \left( \frac{3}{(5c+22)(k+3)^3} \right)^{\frac{1}{2}}
\; W^8,
\ea
generate the $W_3$ algebra with central charge
\be
c(k)=50-24(k+3)-\frac{24}{k+3}  \label{cch}
\ee

For any principal embedding the grade zero subalgebra $\bar{g}_0$ is just
the Cartan subalgebra. In the case at hand $\hat{F}_0$ is therefore a direct
sum of two affine $u(1)$ field algebras, both of level $k+3$, generated
by $\hat{J}^4$ and $\hat{J}^5$.
Defining $\partial \phi_1 \equiv \nu_1 \hat{J}^4$ and $\partial \phi_2
\equiv \nu_2 \hat{J}^5$, where $\nu_1=\sqrt{6(k+3)}$ and $\nu_2=
\sqrt{2(k+3)}$ it is easy to check that
\be
\partial \phi_i (z) \partial \phi_j (w) = \frac{\delta_{ij}}{(z-w)^2}+\ldots
\ee
According to the general prescription the Miura transformation reads in
this case $W^7 \mapsto -W^7_1$ and $W^8 \mapsto W^8_2$, and the fields
\ba
T^{(0,0)} & = & -\nu_2^{-2}W^7_1 \nonumber \\
W^{(0,0)} & = & \left( \frac{3}{(5c+22)(k+3)^3}\right) ^{\frac{1}{2}}W^8_2
\nonumber
\ea
also generate a $W_3$ algebra with central charge (\ref{cch}).
Note that according to (\ref{ttt}) $T^{(0,0)}$ and $W^{(0,0)}$ only
depend on $\phi_1$ and $\phi_2$. This is the well known free field
realization of $W_3$.

\newsubsection{The $W_3^{(2)}$ algebra}

The two examples discussed above are both related to principal $sl_2$
embeddings. In order to illustrate that our methods work for arbitrary
embeddings we now consider the example of the $W_3^{(2)}$ algebra which
is associated to the (only) nonprincipal $sl_2$ embedding into $sl_3$.

To describe the $W_3^{(2)}$ algebra, we pick a slightly
different basis of $sl_3$, namely
\be \label{basis4}
J^at_a=\mats{\frac{J^4}{6}-\frac{J^5}{2}}{J^6}{
J^8}{J^2}{-\frac{J^4}{3}}{J^7}{
J^1}{J^3}{\frac{J^4}{6}+\frac{J^5}{2}}.
\ee
The $sl_2$ embedding reads $t_+= t_1$, $t_0=
t_5$ and $t_- = t_8$. The
gradation of $\bar{g}$ with respect to $ad_{t_0}$ is half-integer
which means that there will be second class constraints \cite{BTV}
corresponding to the fields with grade -1/2. If one wants to use the
BRST formalism all constraints should be first class. One way to get
around this problem is to introduce auxiliary fields \cite{B}. This is not
necessary however as was shown in \cite{FORTW,BT} since it is always
possible to replace the half integer grading and the constraints
associated to it by an integer grading and
a set of first class constraints that nevertheless
lead to the same Drinfeld-Sokolov reduction. As mentioned earlier we have
to replace the grading by $t_0$ by a grading w.r.t. an element $\delta$.
\footnote{Essentially what one does is split the set of second class
constraints in two halves. The constraints in the
first half, corresponding to positive
grades w.r.t. $\delta$, are still imposed but have now become first class.
The other half can then be obtained as gauge fixing conditions of the
gauge invariance generated by the first half.}
In this specific case  $\delta = \frac{1}{3}\mbox{diag}(1,1,-3)$.
With respect to $ad_{\delta}$,
$t_1$ and $t_3$ have degree $1$ and span $\bar{g}_+$. The BRST
current is
\be \label{brst32}
d(z)=((J^1-1)c_1)+J^3c_3.
\ee
Notice that there is no need for auxiliary fields, since the
constraints $J^1=1$ and $J^3=0$ are first class. The
cohomology $H(F_{red}(\Omega_k);D_0)$ is generated by $\{\hj^4,
\hj^6,\hj^7,\hj^8\}$. Again using the tic-tac-toe construction one finds
that $H(F(\Omega_k);D)$ is generated by $W^4=W^4_0$; $W^6=W^6_0$;
$W^7=W^7_0-W^7_1$ and $W^8=W^8_0-W^8_1$, where
\ba
W^4_0 & = & \hj^4, \nonu
W^6_0 & = & \hj^6, \nonu
W^7_0 & = & \hj^7, \nonu
W^7_1 & = & \frac{1}{2}(\hj^2\hj^5) + \frac{1}{2}(\hj^2\hj^4) -
(k+1) \dif \hj^2, \nonu
W^8_0 & = & \hj^8, \nonu
W^8_1 & = & -\frac{1}{4} (\hj^5 \hj^5) - (\hj^2\hj^6) +
\frac{(k+1)}{2}\dif\hj^5.
\ea
The OPEs of the hatted currents involving shifts in level are in this case
\ba \label{hjope}
\hj^4(z)\hj^4(w) & \sim & \frac{6(k+\frac{3}{2})}{(z-w)^2} + \cdots \nonu
\hj^5(z)\hj^5(w) & \sim & \frac{2(k+\frac{5}{2})}{(z-w)^2} + \cdots \nonu
\hj^4(z)\hj^5(w) & \sim & \frac{3}{(z-w)^2}+\cdots \nonu
\hj^2(z)\hj^6(w) & \sim & \frac{(k+1)}{(z-w)^2} +
\frac{\frac{1}{2}(\hj^4-\hj^5)}{(z-w)} +
\cdots.
\ea
If we now define the following generators
\ba
H & = & -W^4/3, \nonu
G^+ & = & W^6, \nonu
G^- & = & W^7, \nonu
T & = & \frac{1}{k+3}(W^8+\frac{1}{12}(W^4 W^4)),
\ea
we find that these generate the $W_3^{(2)}$ algebra
\cite{B}, with
\be
c(k)=25-6(k+3)-\frac{24}{k+3}
\ee
a formula that was found in \cite{B} by a counting argument.

In the case at hand the subalgebra $\bar{g}_0$ is spanned by the elements
$t_4,\; t_5,\; t_6$ and $t_2$. Obviously $\bar{g}_0 \simeq sl_2 \oplus u(1)$.
Therefore $\hat{F}_0$ is the direct sum of an affine $sl_2$ and an affine
$u(1)$ field algebra, and using equation (\ref{shift}) the levels of these
can be calculated to be $k+1$ and $k+3$ respectively.
Indeed if we introduce the currents
\ba
\partial \phi & = & \frac{1}{4} (\hat{J}^4+3\hat{J}^5) \nonumber \\
J^0 & = & \frac{1}{4} (\hat{J}^5 - \hat{J}^4) \nonumber \\
J^- & = & \frac{1}{2} \; \hat{J}^2 \nonumber \\
J^+ & = & 2 \; \hat{J}^6 \nonumber
\ea
then these satisfy the following OPEs
\ba
J^0(z)J^{\pm}(w) & = & \frac{\pm J^{\pm}(w)}{z-w}+\ldots
\nonumber \\
J^+(z)J^-(w) & = & \frac{(k+1)}{(z-w)^2}+\frac{2J^0(w)}{z-w}+
\ldots \nonumber \\
J^0(z)J^0(w) & = & \frac{\frac{1}{2}(k+1)}{(z-w)^2}+\ldots
\nonumber \\
\partial \phi (z)\, \partial \phi (w) & = & \frac{\frac{3}{2}(k+3)}{
(z-w)^2}+ \ldots  \nonumber
\ea
and all other OPEs are regular. As stated before the shifts in the
levels that one can see in these OPEs are a result of the ghost
contributions.

The quantum Miura transformation in this case reads: $W^4 \mapsto W^4_0,\;
W^6 \mapsto W^6_0,$ \\ $W^7 \mapsto -W^7_1, \; W^8 \mapsto -W^8_1$. This means
that in terms of the currents $J^{\pm},\; J^0$ and $\phi$ the grade (0,0)
components of the $W$ generators read (let's for notational convenience
denote the (0,0) components of $H,G^+,G^-$ and $T$ again simply by
the same letters since they generate an isomorphic algebra anyway)
\ba
H & = & J^0-\frac{1}{3}\partial \phi \nonumber \\
G^+ & = & \frac{1}{2}J^+ \nonumber \\
G^- & = & (J^-J^0)+(J^0J^-)+2(k+3)\partial J^--2 \partial \phi J^-
\nonumber \\
T & = & \frac{1}{2(k+3)} \left( 2(J^0J^0)+(J^-J^+)+
(J^+J^-) +(k+3)\partial J^0
+ \frac{2}{3}(\partial \phi \partial \phi ) +
Q_0\partial^2 \phi \right) \nonumber
\ea
where $Q_0=-(k+1)$. We recognise
in the expression for $T$ the improved $sl_2$ Sugawara stress energy tensor
and the free boson in a background charge.
Note that these formulas provide an embedding
of $W_3^{(2)}$ into $\hat{F}_0$.
In \cite{BTV} a realization of this type was called a
'hybrid field realization' since it represents the $W$ algebra partly
in terms of KM currents and partly in terms of free fields.

It is
now easy to obtain a realization of $W_3^{(2)}$ completely in terms
of free fields by inserting for the $sl_2$ KM currents $J^{\pm},\; J^0$
their Wakimoto free field form
\ba
J^- & = & \beta \nonumber \\
J^+ & = & -(\gamma^2 \beta )-k\partial \gamma - \sqrt{2k+6}
\,\gamma \,i \partial \varphi
\nonumber \\
J^0 & = & -2(\gamma \beta )-\sqrt{2k+6} \,i \,\partial \varphi
\ea
where as usual $\beta, \gamma$ and $\varphi$ are bosonic fields with the
following OPEs
\ba
\gamma (z) \beta (w) & = & \frac{1}{z-w}+ \ldots \nonumber \\
i\partial \varphi (z)\, i \partial \varphi (w) & = & \frac{1}{(z-w)^2}+\ldots
\ea

This example gives a nice taste of the general case. By the Miura
transformation one can write down for any $W$ algebra a hybrid field
realization, i.e. a realization partly in terms of free fields and partly
in Kac-Moody currents. When required one can then insert for the KM currents
the Wakimoto free field realization giving you a realization of the $W$
algebra completely in terms of free fields.

\newpage

\newsection{Discussion}
In this paper we have quantized all generalized Drinfeld-Sokolov
reductions. This was done using a formalism that differs from the
formalism first used by Bershadsky and Ooguri. The formalism of
Bershadsky-Ooguri makes use of the Fock space resolutions of
affine Lie algebras and $W$ algebras. In the calculation of the BRST
cohomology
they have to prove that the BRST cohomology and the resolution
cohomology commute. This they indeed did for the Virasoro algebra
but in the case of the $W_3^{(2)}$ algebra it is an assumption \cite{B}.
The $W$ algebras are in the end constructed as the commutant of
certain screening operators. Calculating this commutant and finding
a complete set of generators of it is in general very difficult. In
\cite{B} Bershadsky doesn't prove that the generators that he
provides are a complete set nor does he show how he has
obtained them.
Our method on the other hand does not make any assumptions, is
completely algorithmic and works for arbitrary $sl_2$ embeddings. The
difference with the cohomology calculations of Feigin and
Frenkel \cite{FeFr} is that the spectral sequence they use is different
from the one that we use (in principal one can associate two spectral
sequences to any double complex).

An important open problem is to find unitary representations of
the $W$ algebras in the set $\cal W$. It is believed that many
questions about the representation theory can be answered using
the correspondence between Lie algebras and $W$ algebras exhibited
in this paper. For example it should be possible to derive character
formulas for the $W$ algebras from the affine characters (see also
\cite{FKW}). This is now under investigation.

\newsection{Appendix}

In this appendix we review some basic facts on $sl_2$ embeddings \cite{Dynkin}.
The $sl_2$ embeddings into $sl_n$ are in one to one correspondence with
the partitions of $n$. Let $(n_1,n_2 \ldots )$ be a partition of $n$ with
$n_1 \geq n_2 \geq \ldots$ then define a different partition $(m_1, m_2,
\ldots )$ of $n$ by letting $m_k$ be the number of $i$ for which $n_i\geq
k$. Furthermore let $s_t=\sum_{i=1}^t m_i$. Then the $sl_2$ embedding
associated to the partition $(n_1,n_2,\ldots )$ is given by
\ba
t_+ & = & \sum_{l\geq 1}\sum_{k=1}^{n_l-1}E_{l+s_{k-1},l+s_k}, \nonumber \\
t_0 & = & \sum_{l\geq 1}\sum_{k=1}^{n_l}(\frac{n_l+1}{2}-k)
E_{l+s_{k-1},l+s_{k-1}}, \nonumber \\
t_- & = & \sum_{l\geq 1}\sum_{k=1}^{n_l-1}k(n_l-k)E_{l+s_k,l+s_{k-1}}
\nonumber
\ea
where $E_{ij}$ is as usual the $n \times n$ matrix with zeros everywhere
except for the matrix element $(i,j)$  which is equal to one.
The element $\delta$ which defines the grading on $sl_n$ that we use to
impose the constraints is given by \cite{BT}
\be
\delta = \sum_{k\geq 1}\sum_{j=1}^{m_k} \left( \frac{\sum_l lm_l}{\sum_l
m_l}-k \right) E_{s_{k-1}+j,s_{k-1}+j}.
\ee
One can check that in case the grading provided by $t_0$ is integer
then $\delta = t_0$.

The fundamental representation  of $sl_n$ decomposes into
irreducible $sl_2$ multiplets. This we denote by $\underline{n}\rightarrow
\oplus_i n_i \underline{i}$, where $\underline{i}$ is the
$i$-dimensional representation of $sl_2$. We then have the following
identities that come in useful when trying to calculate the central
charge $c(k;\delta )$ for a certain specific case.
\ba \label{someid}
\frac{1}{2}\dim(\bar{g}_{\frac{1}{2}}) & = &
\sum_{i>0,k\geq 0} in_in_{i+2k+1}, \nonu
|\rho|^2 & = & \frac{1}{12} (n^3-n), \nonu
|t_0|^2 & = & \frac{1}{12} \sum_i n_i (i^3-i), \nonu
(t_0|\rho) & = & \frac{1}{12}\left(\sum_i n_i^2 (i^3-i) +
\sum_{i<r} i(i+1)(3r-i-2)n_in_r \right).
\ea
This concludes our discussion of $sl_2$ embeddings.

\newsection{Acknowledgements}
We would like to thank J. Goeree and F.A.Bais for useful discussions
and comments.

\end{document}